\documentstyle[12pt,a4,epsfig]{article}

\newcommand{\gap}{\stackrel{>}{\sim}}

\begin{document}
\sloppy
\thispagestyle{empty}

\rightline{MSU-51120}

\mbox{}
\vspace*{\fill}
\begin{center}
{\LARGE\bf A Collection of Polarized Parton Densities
\footnote{Presented during the Workshop on the {\it Prospects of
Spin Physics at HERA} held at DESY-Zeuthen, 
Germany, 28-31 August 1995.}}
 \\

\vspace{2em}
\large
Glenn A. Ladinsky
\\
\vspace{2em}
{\it  Michigan State University}
 \\
{\it Department of Physics \& Astronomy, East Lansing, MI  48824-1116
 U.S.A.}\\
\end{center}
\vspace*{\fill}
\begin{abstract}
\noindent
A significant number of parameterizations for the polarized parton densities
have appeared in the literature.  Using the CTEQ evolution package,
these distributions have been evolved consistently preparatory to
compilation into an integrated package
in the spirit of PDFLIB by Plothow-Besch.
Here, a comparison of a few of the more recent distributions are made.
\end{abstract}
%
\def\journal#1&#2(#3)#4{{\unskip,~\sl #1\unskip~\bf\ignorespaces #2\unskip~\rm (19#3) #4}}

\label{sect1}
\section{Introduction}

The ideas for computing polarized parton densities have been around for
a long time\cite{kaur}.  What has been lacking is sufficient experimental data 
to define those densities.  
Good progress has been made from deep inelastic scattering experiements.
At present, the data ranges are from $0.003<x<0.8$  
and $1\,$GeV${}^2<Q^2<60\,$GeV${}^2$,
and this has provided reasonable fits descibing the up quark and
down quark distributions.  Data intended to constrain the gluon
and sea quark densities tightly has yet to be obtained.  

Recent theoretical progress has given us the next to leading order
Altarelli-Parisi splitting kernels for polarized partons\cite{rolf}.  
The first helicity distributions based on this higher order evolution 
have already appeared\cite{grsv}.  

\begin{table}[t]
\caption{This is a partial listing of parameterizations for
the polarized parton distribution functions.}
\begin{center}
\begin{tabular}{ l | l | l }
\hline\hline
\multicolumn{3}{c}{$\Delta$PDFs}\\
\hline\hline
$\Delta$PDFs & AUTHORS  & { REFERENCE} \\
\hline
BT-95  & Bartelski \& Tatur            & { preprint, CAMK 95-288} \\
GRSV-95& Gluck, Reya, Stratmann \& Vogelsang
                                       & { preprint, DO-TH 95/13} \\
GRV-95 & Gluck, Reya \& Vogelsang      & { preprint, DO-TH 95/11} \\
CLW-95 & Cheng, Liu and Wu             & { preprint, IP-ASTP-17-95} \\
BS-95  & Bourrely and Soffer           & { Nucl.~Phys.} {B445} (1995) 341 \\
F-95   & de Florian, et al.            & { Phys.~Rev.} {D51} (1995) 37 \\
GS-95  & Gehrmann and Stirling         & { Z.~Phys.} {C65} (1995) 461 \\
BBS-95 & Brodsky, Burkardt \& Schmidt  & { Nucl.~Phys.} {B441} (1995) 197 \\
N-94   & Nadolsky                      & { Z.~Phys.} {C63} (1994) 601 \\
CCGN-93& Chiappetta, et al.            & { Z.~Phys.} {C59} (1993) 629 \\
F-93   & de Florian, et al.            & { Phys.~Lett.} {B319} (1993) 285 \\
CW-92  & Cheng and Wai                 & { Phys.~Rev.} {D46} (1992) 125 \\
SL-92  & Sridhar and Leader            & { Phys.~Lett.} {B295} (1992) 283 \\
CN-91  & Chiappetta and Nardulli       & { Z.~Phys.} {C51} (1991) 435 \\
GRV-90 & Gluck, Reya and Vogelsang     & { Nucl.~Phys.} {B329} (1990) 347 \\
G-90   & Gupta, et al.                 & { Z.~Phys.} {C46} (1990) 111 \\
CL-90  & Cheng and Lai                 & { Phys.~Rev.} {D41} (1990) 91 \\
\hline\hline
\end{tabular}
\end{center}
\end{table}

A partial listing of polarized parton densities is given in Table~1.
As we can see, 1995 has been a good year.  Soon, we should have about
as many distributions as there are data points to fit.
This table mainly focuses on those distributions from the 1990's, but
there are other distributions that people have found useful that
appeared before 1990\cite{extras}.

\begin{table}[t]
\caption{This is a listing of input parameters for the polarized
parton distribution functions in Table~1.}
\begin{center}
\begin{tabular}{ l | l | l | c | c | l}
\hline\hline
\multicolumn{6}{c}{$\Delta$PDFs}\\
\hline\hline
Mode & $\Delta$PDF  & Order & $Q^2_0$ (GeV${}^2$) & 
              $\Lambda_{QCD}^{(4)}$ (MeV${}^2$)& Unpolarized PDF\\
\hline
270 & BT-95    & LO  &  4.0 & 230 & U-MRSA (set A)\\
260 & CLW-95   & LO  & 10.0 & 231 & U-MRSA${}^\prime$ (set A${}^\prime$)\\
250 & GRSV-95  & NLO &  0.34& 200 & U-GRVt-95 \\
240 & GRV-95   & LO  &  0.23& 200 & U-GRVt-95 \\
230 & BS-95    & LO  &  3.0 & 200 & BS-95 \\
220 & F-95     & LO  & 10.0 & 230 & U-MRS${}^\prime$ (set D${}_-^\prime$) \\
210 & GS-95    & LO  &  4.0 & 177 & U-O \\
200 & BBS-95   & LO  &  4.0 & 230 & U-MRS${}^\prime$ (set D${}_0^\prime$) \\
190 & N-94     & LO  & 11.0 & 200 & U-GRVt-92 \\
180 & CCGN-93  & LO  &  1.0 & 260 & U-DFLM (avg. set) \\
170 & F-93     & LO  &  4.0 & 168 & U-CTEQ1 \\
160 & CW-92    & LO  & 10.0 & 260 & U-DFLM (avg. set) \\
150 & SL-92    & LO  &  4.0 & 177 & U-O \\
140 & CN-91    & LO  &  1.0 & 260 & U-DFLM (avg. set) \\
130 & GRV-90   & LO  & 10.0 & 360 & U-GHR \\
120 & GPS-90   & LO  &  5.0(15.0) & 200(90) & U-EHLQ(U-EMC) \\
110 & CL-90    & LO  & 10.7 & 260 & U-DFLM (avg. set)\\
\hline\hline
\end{tabular}
\end{center}
\end{table}

Using the CTEQ evolution package\cite{cteq}, these distributions have
been evolved consistently to allow for a comparison of a few of the
more recent distributions as well as to facilitate future research.  A
library in the spirit of PDFLIB by Plothow-Besch\cite{pdflib} is under
development and will soon be available for distribution.  In the
remainder of this report, details of the evolution are presented and
some comparisons between a few distributions from the 1990's are made.

\label{sect2}
\section{On the Parameters of the $Q^2$ Evolution}

To have a properly defined parton distribution function (PDF) requires that
a number of parameters be defined from the outset (something about which many
papers are not very explicit).  In Table~2 the values of the
parameters used in this evolution are presented.  
Generally, the evolution for each parton
distribution starts with an initial distribution at a given energy scale
($Q_0$) as provided in the published work. Besides the initial scale for
the evolution and the order in perturbation theory in which the
evolution is being performed, it is also necessary to set the QCD scale
($\Lambda_{QCD}^{(n_f)}$) for a given number of quark flavors
$n_f$ and the quark masses.  
For those papers that did not explicitly
present a value for $\Lambda_{QCD}^{(n_f)}$, the value was taken from
their choice of unpolarized PDF.

The quark masses are not mentioned in the tables.  A few of the
distributions begin their evolution at scales of $Q_0$ around 1~GeV,
which can be below the charm and bottom quark thresholds.  This has
its effect on the evolution when these quark thresholds are crossed.
Some authors indicate what quark masses were used, like in
GRV-95 where $m_c=1.5\,$GeV and $m_b=4.5\,$GeV were given, but for
most of the cases values of $m_c=1.6\,$GeV and $m_b=5.0\,$GeV
have been adopted.
In all the evolution
performed here, the top quark mass has been set to $m_t=180\,$GeV.

Since the data has been at low $Q^2$, quite a number of papers perform
their fits based on evolutions with three quark flavors.  For the
evolutions here, however, the full six flavors are used with the
understanding that some people want to do computations at higher
energy scales.

\begin{table}[t]
\caption{This is a listing of references for the unpolarized parton 
distribution functions associated with the $\Delta$PDFs in Table~1.}
\begin{center}
\begin{tabular}{ l | l | l }
\hline\hline
\multicolumn{3}{c}{Unpolarized PDFs}\\
\hline\hline
PDFs & AUTHORS  & { REFERENCE} \\
\hline
U-GRVt-95& Gluck, Reya and Vogt         & { Z.~Phys.} {C67} (1995) 433 \\
U-BS     & Bourrely and Soffer          & { Nucl.~Phys.} {B445} (1995) 341 \\
U-BBS    & Brodsky, Burkardt \& Schmidt & { Nucl.~Phys.} {B441} (1995) 197 \\
U-MRSA${}^\prime$  
         & Martin, Stirling \& Roberts  & { Phys.~Lett.} {B354} (1995) 155 \\
U-MRSA   & Martin, Roberts \& Stirling  & { Phys.~Rev.} {D56} (1994) 6734 \\
U-MRS${}^\prime$   
         & Martin, Roberts \& Stirling  & { Phys.~Lett.} {B306} (1993) 145 \\
U-CTEQ1  & CTEQ Collaboration           & { Phys.~Lett.} {B304} (1993) 159  \\
U-GRVt-92& Gluck, Reya \& Vogt          & { Z.~Phys.} {C53} (1992) 127 \\
U-O      & Owens                        & { Phys.~Lett.} {B266} (1991) 126 \\
U-DFLM   & Diemoz, et al.               & { Z.~Phys.} {C39} (1988) 21 \\
U-EMC    & Sloan, Smajda and Voss       & { Phys.~Rep.} {162} (1988) 45 \\
U-DO     & Owens                        & { Phys.~Rev.} {D30} (1984) 49 \\
U-EHLQ   & Eichten, et al.              & { Rev.~Mod.~Phys.} {56} (1984) 579 \\
U-GHR    & Gluck, Hoffman \& Reya       & { Z.~Phys.} {C13} (1982) 119 \\
\hline\hline
\end{tabular}
\end{center}
\end{table}

Each set of helicity distributions has been associated in some manner
with a specific set of unpolarized PDFs.  In most cases, this appears
through the application of a dilution model, whereby the $\Delta$PDF
is written as a linear combination of unpolarized PDFs weighted by a
phenomenological function in $x$\cite{kaur}. This association between
the $\Delta$PDFs and the unpolarized PDFs is tabulated in Table~3.  In
performing the evolutions, these associations have been maintained.

The $\Delta$PDFs that have been chosen for presentation are
N-94 (sets 1 and 2), BBS-95, GS-95 (sets A, B, and C), and
GRV-95 (``standard'' and ``valence'' scenarios).
In Figs.~1-3, these distributions are shown at the scale $Q=15\,$GeV.
To evolve the BBS-95 helicity densities, it was necessary to 
assume a form for the sea quark distributions.  Since the BBS-95 distributions
are close to the MRSD0${}^\prime$ form, those sea quark densities were used to
provide the helicity densities for the sea; namely, at the initial
energy scale it was assumed that $\Delta\bar{u}(x)=\bar{u}(x)/2$ and 
$\Delta\bar{d}(x)=\bar{d}(x)/2$.

\label{sect3}
\section{The Up and Down Quarks}

The two distributions best defined by the present experimental data are the 
up and down quark densities.  Looking at Fig.~1, there is nice agreement
between the different fits for $\Delta u$; it isn't until the larger
$x$ are reached (where the error bars on the data increase) that 
significant deviations occur.  The down quark densities appear
to have a few atypical contenders, but it should be noted that in
the plot of $x\Delta d(x)$, the BBS-95 and N-94 (set 2) distributions
both allow $\Delta d(x)$ to cross over into positive values at some $x$.
Since the $\Delta$PDFs are constrained by the moments, 
where $\Delta d=\int \Delta d(x)\,dx<0$, the BBS-95 
and N-94 (set 2) distributions compensate
for their positive contribution to the integral with a more negative
$\Delta d(x)$ below the crossover point.

\label{sect4}
\section{The Gluon and Sea Quark Densities}

Since the gluon and sea quark densities are, to a large degree, unconstrained,
it is here where the parameterizations are most distinguishable. 
Some models have negligible or large $\Delta G(x)$ and $\Delta s(x)$
while others may carry more moderate $\Delta G(x)$ and $\Delta s(x)$.
Hopefully, this issue will be settled with results from hadron-hadron
collisions at laboratories like the RHIC.\cite{rhic}

\label{sect5}
\section{Large $x$ Limits}

As discussed by Farrar and Jackson\cite{farrarj}, 
our expectation is that as the momentum fraction of the parton approaches
unity that the helicity of the parton should coincide with that
of its parent hadron.  In other words, we have the limit
\begin{equation}
\Delta{q}(x)/q(x) \longrightarrow 1\qquad \hbox {as}\qquad {x\rightarrow 1}.
\label{eq:fjlimit}
\end{equation}

The plots in Fig.~2 illustrate the polarization of some of the $\Delta$PDFs
in the large $x$ limit.  The BBS-95 distributions carry a smooth transition
of the polarization towards unity with large $x$ while other distributions
have polarizations that sharply rise at very large $x$,
plateau around $x\gap 0.6$, or ignore this large $x$ behavior.
Numerically, the PDFs are small as $x$ nears unity, which minimizes the
effect such variations in the $\Delta$PDF may have on physics.
Nevertheless, it is important to note that it is in the large $x$ region
that the polarization distinguishes between the different models and
fits\cite{nadw}.  In particular, different parameterizations of the
$\Delta d$ distribution cross over from negative to positive values
at different $x$.

\label{sect6}
\section{Small $x$ Extrapolations}

With the larger energy colliders like HERA, the RHIC or the LHC
comes the possibility of invesitgating polarization physics at higher energy
scales and smaller momentum fraction than fixed target experiments.  
In Fig.~3, the small-$x$ extrapolation
for a sample of $\Delta$PDFs is displayed.  These results indicate that,
as usual, care must be exercised when extending the use of the $\Delta$PDFs
beyond the range of the data with which they were fit.  The view generally
taken is that the polarization of the helicity distributions should
vanish as $x\rightarrow 0$\cite{brodsky}.  Using the distributions beyond
their range of validity can produce unreasonable results.

\label{sect7}
\section{$Q^2$ Evolution}

In Fig.~4 the $Q^2$ evolution is performed for four of the $\Delta$PDFs
we have been examining.  What is specifically shown is 
the charge weighted sum over the quark helicity distributions,
\begin{equation}
{1\over 2}\sum_i e^2_i \Delta f_i(x,Q^2),
\label{eq:fisum}
\end{equation}
for $Q=0.015,0.1,1,10\,$TeV and
where $i$ runs over the quark types $u,d,s,\bar{u},\bar{d},\bar{s}$.
(I do not call this $g_1$ because some papers, such as GS-95,
define this structure
function differently by including the anomalous gluon contribution.)
The thing to note in Fig.~4 is that the $Q^2$ evolution 
reveals distinguishing features of the different helicity
distributions, indicating that the evolution properties themselves
will be useful to consider when establishing the $\Delta$PDFs\cite{blum}.

\label{sect8}
\section{Caveats}

This work has mainly been a presentation of polarized parton densities
available in the literature.  Though some features of these
distributions have been discussed, no attempt has been made here to
determine the quality or correctness of the fits or the procedures
used.  In some cases the comparison would be pointless because of the
improvements in data and theory over the past decade.

There are a few caveats for the user. The many distributions have been
fit with different data at different times and may eventually become
outdated.  Furthermore, not all the distributions have been determined
from the same perspective within QCD.  An example of such variations
can be found in the different definitions used for the structure
function $g_1$.  Inconsistencies, many of which are irrelevant due to
the lack of constraining data on the gluon and sea quark densities,
may also appear in the fits ({\it e.g.}, higher order unpolarized PDFs
sometimes have been used as the input distribution for helicity
densities evolved in leading order).  Nonetheless, with the variety
available among all the $\Delta$PDFs, a wide range of possible physics
can be investigated.

\label{sect9}
\section*{Acknowledgments}

My gratitude goes to Wolf-Dieter Nowak and Johannes Blumlein
for inviting me to DESY-Zeuthen and for their hospitality
during my stay.  Thanks also go to Wu-Ki Tung for the
CTEQ routines.
This work is funded in part by DESY-Zeuthen,
Michigan State University and NSF grant PHY-9309902.


\normalsize

\newpage
\begin{center}
\mbox{\epsfig{file=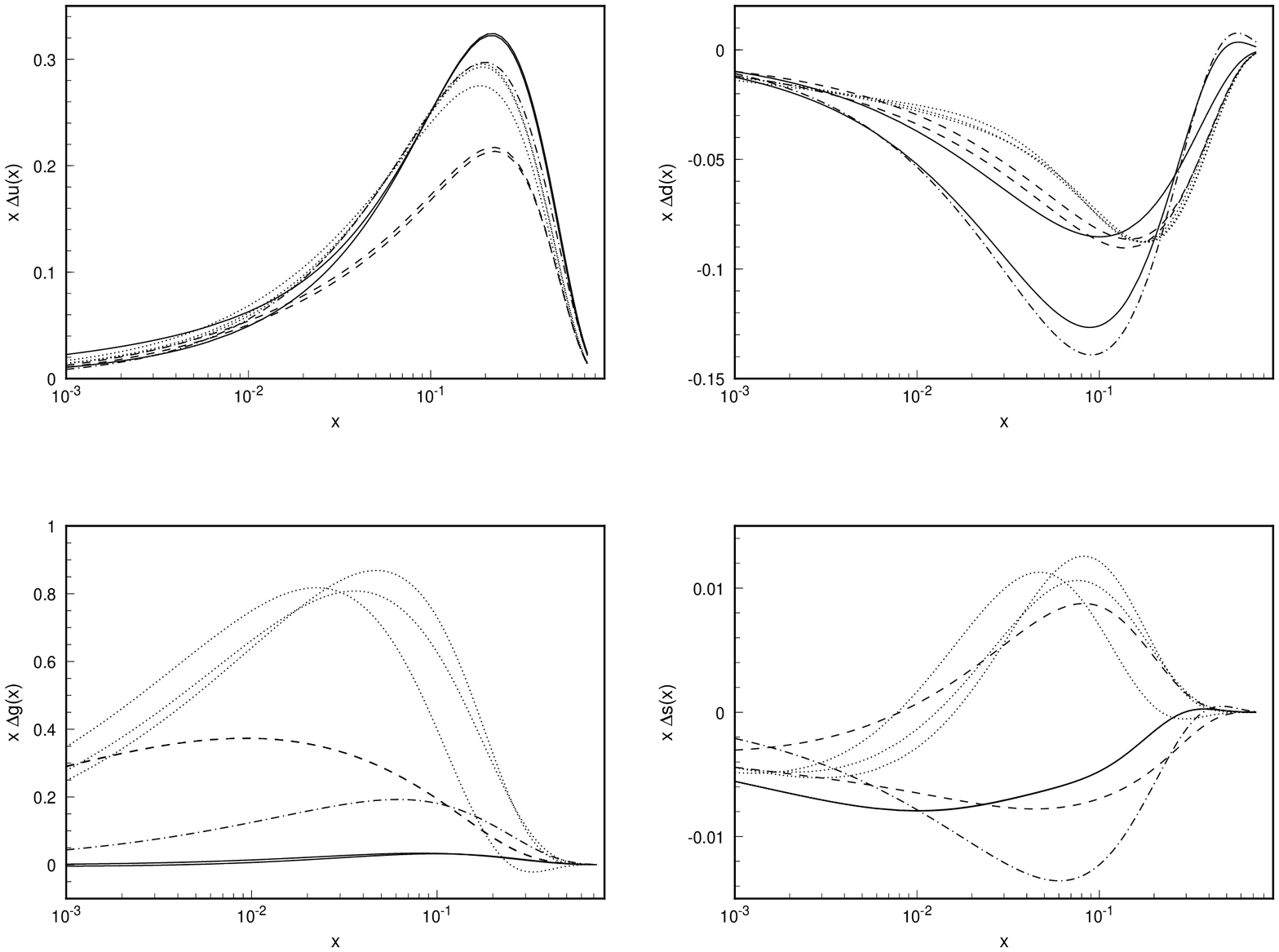,height=18cm,width=18cm}}
\vspace{2mm}
\noindent
\small
\end{center}
{\sf Figure~1:}~The helicity distributions for up, down and strange quarks
and for gluons are shown at $Q=15\,$GeV:
N-94 sets 1 and 2 (solid lines), BBS-95 (dash-dot), 
GS-95 sets A, B, and C (dotted), and
the GRV-95 ``standard'' and ``valence'' scenarios (dashed).

\begin{center}
\mbox{\epsfig{file=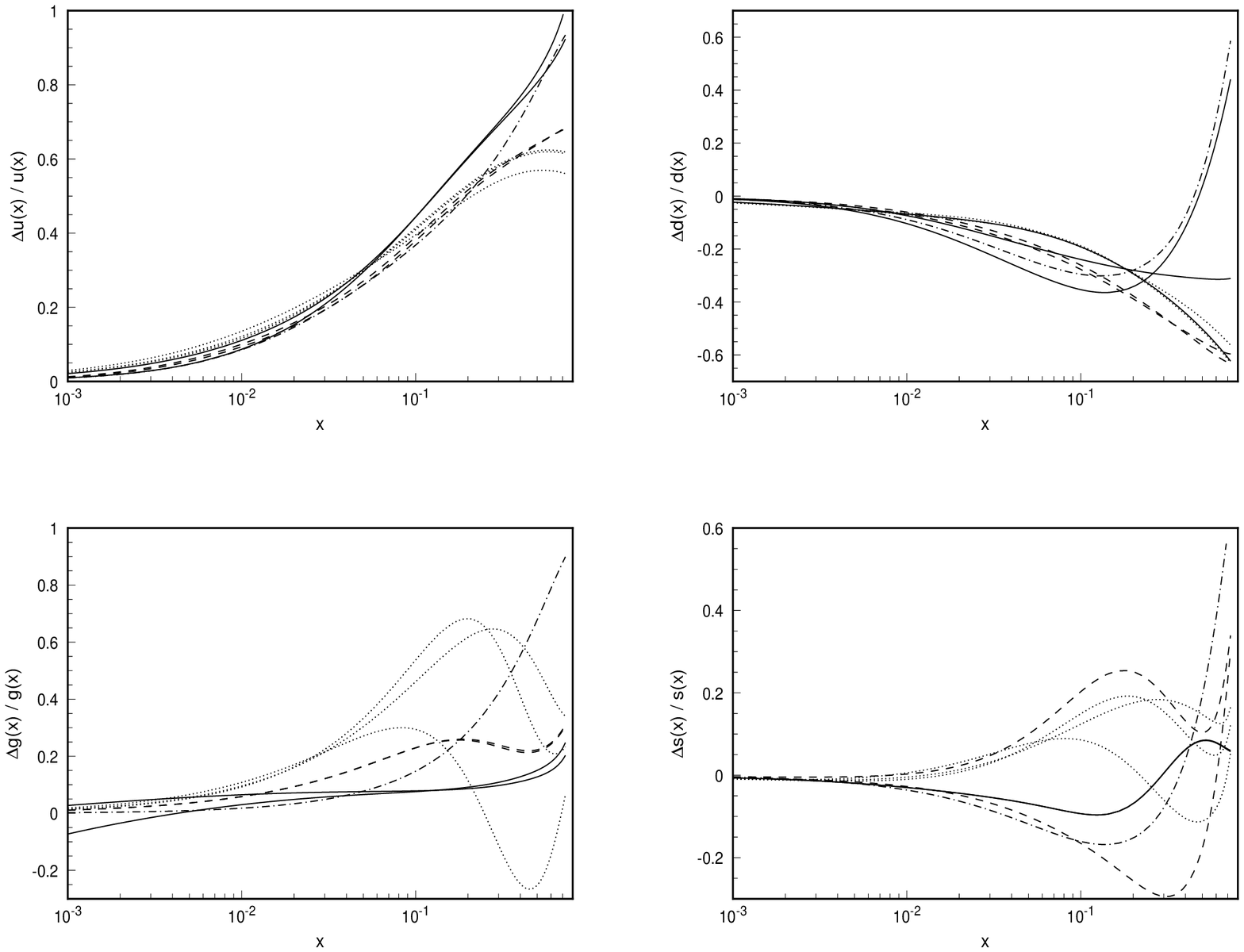,height=18cm,width=18cm}}
\vspace{2mm}
\noindent
\small
\end{center}
{\sf Figure~2:}~The polarization of the partons in the proton as described
by the ratios $\Delta u(x)/u(x)$, $\Delta d(x)/d(x)$, $\Delta g(x)/g(x)$
and $\Delta s(x)/s(x)$.  Curves are as in Fig.~1.

\begin{center}
\mbox{\epsfig{file=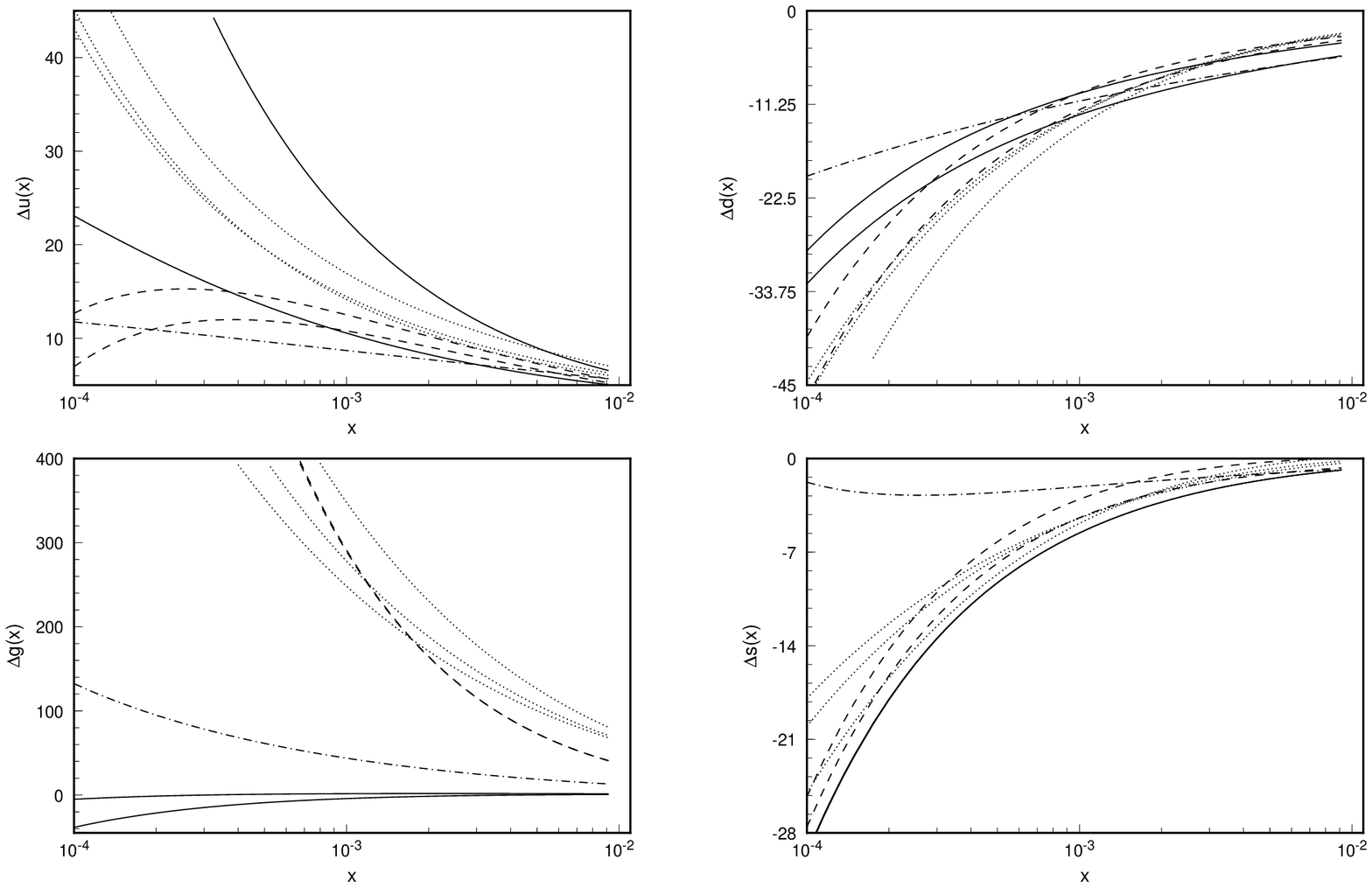,height=18cm,width=18cm}}
\vspace{2mm}
\noindent
\small
\end{center}
{\sf Figure~3:}~The small-$x$ extrapolations for the
helicity distributions of the up, down and strange quarks
and for the gluons are shown.  Curves are as in Fig.~1.

\begin{center}
\mbox{\epsfig{file=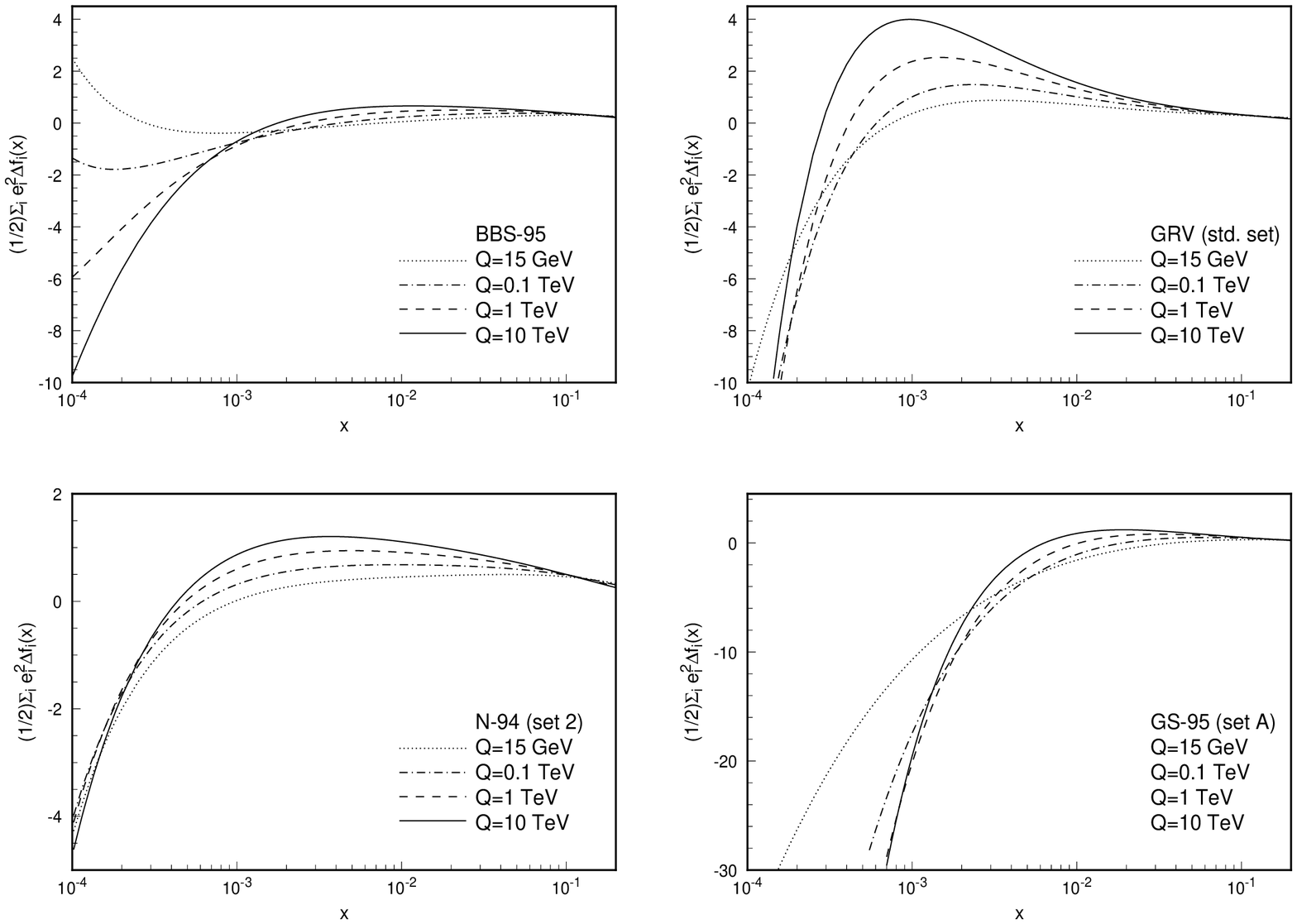,height=18cm,width=18cm}}
\vspace{2mm}
\noindent
\small
\end{center}
{\sf Figure~4:}~The $Q^2$ evolution for four of the $\Delta$PDFs 
are displayed.

\normalsize
\end{document}